# Towards Generalized Parameter Tuning in Coherent Ising Machines: A Portfolio-Based Approach

Tatsuro Hanyu
Graduate School of Informatics,
Nagoya University
Aichi, Japan
hanyu@hpc.itc.nagoya-u.ac.jp

Takahiro Katagiri
Information Technology Center,
Nagoya University
Aichi, Japan
katagiri@cc.nagoya-u.ac.jp

Daichi Mukunoki
Information Technology Center,
Nagoya University
Aichi, Japan
mukunoki@cc.nagoya-u.ac.jp

Tetsuya Hoshino
Information Technology Center,
Nagoya University
Aichi, Japan
hosino@cc.nagoya-u.ac.jp

*Abstract*— Coherent Ising Machines (CIMs) have recently gained attention as a promising computing model for solving combinatorial optimization problems. In particular, the Chaotic Amplitude Control (CAC) algorithm has demonstrated high solution quality, but its performance is highly sensitive to a large number of hyperparameters, making efficient tuning essential. In this study, we present an algorithm portfolio approach for hyperparameter tuning in CIMs employing Chaotic Amplitude Control with momentum (CACm) algorithm. Our method incorporates multiple search strategies, enabling flexible and effective adaptation to the characteristics of the hyperparameter space. Specifically, we propose two representative tuning methods, Method A and Method B. Method A optimizes each hyperparameter sequentially with a fixed total number of trials, while Method B prioritizes hyperparameters based on initial evaluations before applying Method A in order. Performance evaluations were conducted on the Supercomputer "Flow" at Nagoya University, using planted Wishart instances and Time to Solution (TTS) as the evaluation metric. Compared to the baseline performance with best-known hyperparameters, Method A achieved up to 1.47× improvement, and Method B achieved up to 1.65× improvement. These results demonstrate the effectiveness of the algorithm portfolio approach in enhancing the tuning process for CIMs.

## I. Introduction

### A. Background

As conventional computing approaches face limitations in solving large-scale combinatorial optimization problems, alternative models—such as quantum annealers and hybrid analog-digital systems—have garnered significant interest [1]. These models include both true quantum annealers that harness quantum effects [2] and pseudo-quantum systems inspired by quantum principles, among which the Coherent Ising Machine (CIM) stands out [3].

CIMs emulate Ising spin dynamics to efficiently address NP-hard and NP-complete problems. A notable variant, the Chaotic Amplitude Control (CAC) algorithm, introduces controlled chaotic behavior to mitigate premature convergence to local minima. Although CAC has demonstrated high solution accuracy, its performance is highly sensitive to numerous hyperparameters, making robust and efficient tuning crucial [4].

Due to the complexity and potential interdependencies among these parameters, relying on a single optimization algorithm may not provide sufficient adaptability across diverse problem settings. This study therefore proposes an algorithm portfolio approach that enables flexible selection among multiple search strategies, aiming to improve the overall effectiveness and efficiency of hyperparameter tuning in CIMs.

### B. Problem of Conventional Methods

Conventional methods for hyperparameter optimization present several limitations, particularly when applied to complex, high-dimensional search spaces.

Commonly used techniques include grid search, random search, Bayesian optimization, evolutionary algorithms, and the Nelder–Mead method [5]. Among these, Bayesian optimization is widely recognized as one of the most effective black-box optimization strategies, as it intelligently prioritizes unexplored regions of the search space with high potential for improvement [5].

However, when applied to the simultaneous optimization of a large number of hyperparameters, Bayesian optimization suffers from the curse of dimensionality. As the dimensionality increases, the efficiency of the search process deteriorates significantly, even with Bayesian methods, due to the difficulty of adequately sampling the expanded search space. Consequently, its effectiveness in high-dimensional scenarios is substantially reduced.

To overcome this limitation, we propose a novel approach grounded in an algorithm portfolio framework, which flexibly combines multiple search strategies. This framework is specifically designed to address the challenges of hyperparameter tuning in the CIM-CACm algorithm, where numerous interdependent parameters must be optimized concurrently within the broader context of Coherent Ising Machines (CIMs).

### C. Research Objectives and Contributions

In light of the limitations of the aforementioned conventional methods, this study aims to propose a novel hyperparameter search method tailored for the CIM-CACm algorithm and to evaluate its effectiveness through comparative performance analysis against existing approaches.

In addition, this research seeks to develop a hyperparameter tuning framework that incorporates two key features not considered in conventional methods:

1. A mechanism for reducing the number of simultaneously tuned hyperparameters, thereby shrinking the search space and enhancing the efficiency of Bayesian optimization.
2. A mechanism for selecting and applying different search algorithms dynamically within the tuning process, allowing the framework to adapt to the characteristics of each hyperparameter.

*D. Organization of this paper*

The paper is organized as follows. Section II describes conventional search methods based on Bayesian estimation and proposal algorithm portfolio approach. Section III explains the problems of existing methods and details the proposed approach. Section IV presents the performance evaluation of the proposed methods. Section V introduces related work and clarifies the positioning of this study. Finally, Section VI provides a summary of the findings presented in this paper.

## II. SEARCH STRATEGY BASED ON BAYESIAN ESTIMATION

*A. Optimization of Bayesian Estimation*

Bayesian optimization [5] is a technique used to find the maximum or minimum of a non-trivial function $f(x)$. Instead of directly evaluating the objective function $f(x)$, which can be computationally expensive, Bayesian optimization utilizes a surrogate model to determine the next sampling point. The function used to select this next point is called the acquisition function.

Bayesian optimization performs the search by iteratively repeating the following steps [5]:

1. Construct and update a surrogate model based on the observed data.
2. Compute the acquisition function from the surrogate model, and select the next hyperparameter configuration by maximizing this function.
3. Evaluate the selected hyperparameter configuration.

*B. Surrogate Model*

A surrogate model is a probabilistic model designed to approximate the objective function or to model its probabilistic characteristics, with the goal of selecting the next evaluation point.

Typical surrogate models include the Gaussian Process (GP) and the Tree-structured Parzen Estimator (TPE) [6]. While GP directly estimates the distribution of the objective function, TPE models the distribution of parameters associated with good evaluation values. These methods determine the next search point based on the acquired sample set $D_n$.

*C. Optuna's Optimization Algorithms*

Optuna [7] is an open-source framework for hyperparameter optimization. It supports a variety of optimization algorithms including TPESampler (TPE), GPSampler (GP), SamdomSampler (Random), GridSampler (Grid), and GPSsampler (CMA-ES) [7]. If no specific algorithm is designated, the TPE is used by default.

Grid performs search on a predefined grid, ensuring uniform coverage of the parameter space. However, because it does not learn from past trials, it may become inefficient.

TPE, a type of Bayesian optimization, estimates promising search points based on past trial results and efficiently searches for better solutions.

CMA-ES employs evolutionary computation techniques and is particularly effective for continuous parameter optimization, though it is not suitable for categorical parameters.

GP is also a Bayesian optimization method that uses Gaussian Processes. It is effective even with a small amount of data, but tends to incur high computational cost.

TPE supports a wider range of conditions compared to other samplers. In particular, it is compatible with pruning, multivariate optimization, parameter dependency handling, and constraint-aware optimization, making it well-suited for a wide variety of optimization problems.

*D. The Algorithm Portfolio Approach*

In this study, we propose a novel hyperparameter optimization framework that leverages the aforementioned optimization algorithms using an algorithm portfolio approach.

Fig. 1 illustrates the concept of the algorithm portfolio approach in the context of Optuna's optimization algorithms.

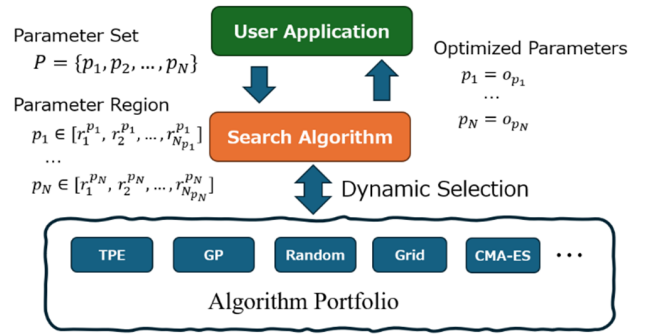

Fig. 1. A concept of the algorithm portfolio approach.

Fig. 1 illustrates a framework in which an algorithm portfolio—a predefined collection of multiple optimization algorithms—is prepared in advance for use within the parameter search process. When performing parameter tuning from a user application, the framework allows dynamic switching among the available algorithms, enabling the selection of the most effective algorithm based on performance during the search.

In the present experiments, we use the optimization algorithms provided by Optuna as the basis for parameter search. However, it should be noted that this is merely one example of implementation; the proposed concept itself is more general and applicable beyond this specific case.

## III. PROPOSAL METHOD

### A. Known Problems for Conventional Searching Methods

Before introducing the proposed methods, we first explain the problems associated with optimizing a large number of parameters simultaneously using Bayesian optimization—issues that arise in conventional approaches.

Fig. 2 shows the result of simultaneously optimizing five hyperparameters of the CIM-CACm algorithm using Bayesian optimization (specifically, the TPESampler) provided by Optuna. The performance is evaluated using Time To Solution (TTS). Details on how TTS is defined and measured are explained in Section IV.C.

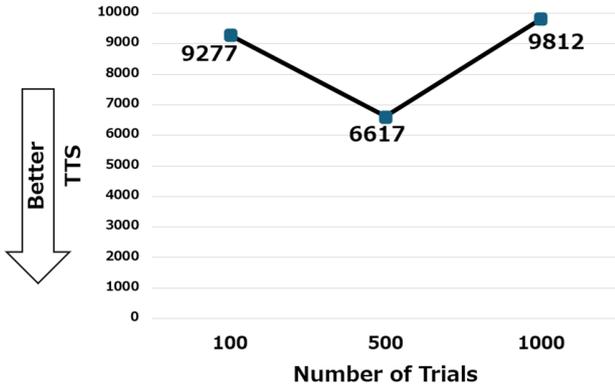

Fig. 2. Problems of conventional Bayesian optimization. The X-axis represents the number of trials, and the Y-axis indicates the TTS (the lower, the better).

From Fig. 2, it can be observed that when the number of hyperparameters being optimized simultaneously is as high as five, increasing the number of trials from 100 to 500 to 1000 does not lead to any noticeable improvement in TTS. This suggests a serious limitation in the optimization method when applied to high-dimensional parameter spaces.

To address this issue, our study attempts to improve performance by reducing the number of hyperparameters explored simultaneously during Bayesian optimization. Specifically, in each optimization run, we limit the search to a single hyperparameter, proceeding with the optimization in a sequential manner.

As will be discussed later, handling cases where multiple hyperparameters simultaneously affect solution quality remains an open challenge for future work.

### B. Proposal Methods

#### 1) Method A

Let the set of hyperparameters be denoted as $P = \{p_1, p_2, \ldots, p_N\}$. An overview of Method A is presented below.

1. Initialization:
   Initialize the parameter set with the initial values $D = \{d_1, d_2, \ldots, d_N\}$, where $d_i \in \mathbb{R}, i = 1, 2, \ldots, N$.
2. Sequential Optimization:
   For each target parameter $p_i$, perform optimization using the hyperparameter optimization function $H$, while keeping all other parameters fixed. In each step, the previously optimized parameters are used as input for the next stage. The number of trials used for optimizing each hyperparameter is set to $T/N$, where $T$ is the total number of trials and $N$ is the number of parameters.
3. Final Evaluation:
   Using the optimized parameter set $P^*$, compute the model evaluation function $M$ to obtain the final evaluation value $E^*$.

Based on this overview, the algorithm for Method A is shown in Algorithm 1.

---
**Algorithm 1** Method A
**Input:** $T$: Total number of trials, $N$: Number of parameters, $P = \{p_1, p_2, \cdots, p_N\}$: Set of parameters, $D = \{d_1, d_2, \cdots, d_N\}$: Set of initial values for parameters, $M(P)$: Evaluation function of model, $H(p, D, T, M)$: Optimization function for hyper parameters.
**Output:** $P^*$: Set of tuned parameters, $E^*$: Final evaluation value.
1: **function** PARAMETER_OPTIMIZATION_A($T, N, P, D, M, H$)
2:    Initialization: $P^{(0)} \leftarrow D$
3:    Set number of trials for each parameter: $T_i \leftarrow T/N$
4:    **for** $i = 1$ to $N$ **do**
5:       Do tuning: $p_i^{(i)} \leftarrow H(p_i, P^{(i)}, T_i, M)$
6:       Update Set of parametes: $P^{(i)} \leftarrow \{P^{(i-1)} \setminus p_i^{(i-1)}\} \cup \{P^{(i)}\}$
7:    **end for**
8:    Optimized set of parameters: $P^* \leftarrow M(P^{(N)})$
9:    Final evaluation value: $E^* \leftarrow M(P^*)$
10:   **return** $P^*, E^*$
11: **end function**

---

Here, $P^{(k)}$ denotes the parameter set at the $k$-th iteration, and $p_i$ is the parameter optimized in that iteration.

In Algorithm 1, note that the tuning method used in line 5 is not restricted to a single optimization algorithm. Furthermore, based on the principle of Method A, if the most suitable search method for the current target parameter $p_i$ is known in advance, it is possible to switch the optimization algorithm accordingly for each parameter during the optimization process.

#### 2) Method B

Method B consists of two stages: an initial evaluation phase and a sequential optimization phase based on Method A. The outline of Method B is described as follows:

1. Initialization:
   Initialize the parameter set $D = \{d_1, d_2, \ldots, d_N\}$ with initial values. Specify the number of trials $T_{initial}$ to be used for the initial evaluation. Then compute the remaining number of trials as $T_{remaining} = T - T_{initial}$.
2. Initial Evaluation:
   For each parameter $p_i$, perform optimization for a fixed number of trials and obtain a temporary optimized result $p_i^{temp}$. Evaluate each result using the model evaluation function $M$ to calculate the evaluation score $E_i$. Based on the evaluation scores $E_i$, sort all parameters in descending order and construct a new parameter sequence $P_{sorted}$.
3. Sequential Optimization:

Based on the prioritized parameter set $P_{sorted}$, call Method A to perform optimization using the remaining number of trials $T_{remaining}$.

4. Final Evaluation:

Obtain the final optimized parameter set $P^*$ and the corresponding final evaluation score $E^*$. The detailed algorithm of Method B is shown in Algorithm 2.

Note that the parameter $T_{initial}$, which specifies the number of trials for initial evaluation, is a performance-sensitive hyperparameter that significantly affects the performance of Method B.

**Algorithm 2** Method B

**Input:** $T$: Total number of trials, $N$: Number of parameters, $P = \{p_1, p_2, \cdots, p_N\}$: Set of parameters, $D = \{d_1, d_2, \cdots, d_N\}$: Set of initial values for parameters, $M(P)$: Evaluation function of model, $H(p, D, T, M)$: Optimization function for hyper parameters, $Y$: Number of trials for initial evaluation.

**Output:** $P^*$: Set of tuned parameters, $E^*$: Final evaluation value.

1: **function** PARAMETER_OPTIMIZATION_B($T, N, P, D, M, H, Y$)
2:    Initialization: $P^{(0)} \leftarrow D$
3:    Set number of trials for initial evaluation: $T_{initial} \leftarrow Y$
4:    Set trial of remainder: $T_{remaining} \leftarrow T - N \times T_{initial}$
5:    #Determine priority of parameters in the initial evaluation.
6:    **for** $i = 1$ to $N$ **do**
7:      Do temporary tuning: $p_i^{temp} \leftarrow H(p_i, P^{(i)}, T_{initial}, M)$
8:      Calculate evaluation value: $E_i \leftarrow M(\{P^{(0)} \setminus p_i^{(0)}\} \cup \{P_i^{temp}\})$
9:    **end for**
10:   Sort based on the calculated evaluation value: $P_{sorted} \leftarrow sort\ (P,\ \text{by}\ E_i\ \text{descending})$
11:   #Do tuning with Method A
12:   Call function: $(P^*, E^*) \leftarrow$ Parameter_Optimization_A $(T_{remaining}, N, P_{sorted}, D, M, H)$
13:   **return** $P^*, E^*$
14: **end function**

As previously noted, in line 12 of Algorithm 2, the internal optimization process in Parameter Optimization A allows for switching between different search algorithms. In addition, Method B also permits switching the search algorithm for each target parameter $p_i$ during the temporary tuning phase in line 7 of Algorithm 2. Moreover, it is possible to evaluate multiple search algorithms during this phase to determine in advance which algorithm yields the best performance. However, increasing the number of trials for this preliminary evaluation may lead to higher computational cost, and thus a trade-off between accuracy and execution time must be considered.

## IV. PEFORMANCE EVALUATION

### A. Coherent Ising Machine (CIM)

In this section, we describe the Coherent Ising Machine (CIM), which serves as the target of our performance evaluation.

Several new types of computing machines have been developed to solve combinatorial optimization problems, and one of them is the CIM [3].

The CIM transforms a combinatorial optimization problem into a problem of finding the minimum-energy state of a theoretical model of interacting spins, known as the Ising model, and solves it experimentally using a physical system that simulates spins. Such machines are often referred to as Ising-type computers [3]. In the Ising model, $N$ spins stabilize in a ground state—the state in which the system's energy is minimized. Finding this ground state is equivalent to finding the solution to the corresponding combinatorial optimization problem. The CIM uses optical parametric oscillators (OPOs) and laser light seeded by incoherent noise (random-phase light) for computation.

There are three main implementations of CIM: (1) actual hardware systems, (2) hybrid systems combining hardware with classical computers, and (3) software simulators running entirely on classical computers. Among these, the simulators are known as Cyber-CIM [8] and Simulated CIM.

### B. Chaotic Amplitude Control with momentum (CACm) method

A known limitation of the CIM is its inability to sufficiently expand the search space. This issue is addressed by the Chaotic Amplitude Control with momentum (CACm) method [3][9], which introduces chaotic behavior into the variables to prevent the CAC from becoming trapped in local minimal.

The parameters used in the CACm are listed in Table I. In addition, the pseudocode for CACm is shown in Fig. 3.

TABLE I.      PARAMETERS OF CACM [10]

| Parameter | Interpretation |
|---|---|
| $T$ | Number of time steps |
| $\beta_1$ | Initial decay rate |
| $\beta_2$ | Final decay rate |
| $\alpha$ | Coupling strength |
| $\gamma$ | Momentum term strength |
| $\xi$ | Rate of change of auxiliary variables |
| $\Delta$ | Time step size |

Fig. 3 illustrates the algorithm for solving an energy minimization problem using the CACm method. In this algorithm, the time evolution of the state variable x(t) and the auxiliary variable e(t) [10] is computed in such a way that the system's energy converges. At the beginning of the algorithm, the damping coefficient β(t) [10] is updated according to the following equation, based on the current time t.

```
FOR t IN 0..T-1
  Set beta to beta1 + t/T*(beta2-beta1)
  Set xpp to xp
  Set xp to x
  Set y to tanh(xp)
  Set mu to w @ y        #Matrix-matrix multiplication
  Set x to xp + Dt*( -beta*xp + alpha*e*mu + gamma*(xp-xpp) )
  Set e to e - (xp**2-1.0)*e*xi
  Set e to e/mean(e)
  Set s to sign(x)
  Set mu0 to w @ s
  Set H to -0.5 sum(s * mu0)
END FOR
```

Fig. 3. Psude-Code of CACm [10].

### C. Coherent Ising Machine with Chaotic Amplitude Control with momentum (CIM-CACm)

The Coherent Ising Machine with Chaotic Amplitude Control with momentum (CIM-CACm) [3][9] is a type of CIM that employs chaotic amplitude control. A Python-based simulator for CIM-CACm is available [10].

CIM-CACm is designed to search for the ground state of the Ising model described by the following Hamiltonian in Equation (1).

$$H(\sigma) = \frac{1}{2}\sigma^T \Omega \sigma, \quad (1)$$

where

$$\sigma \in \{-1, -1\}^N. \quad (2)$$

The descriptions of the variables in Equation (1) and Equation (2) are as follows:
- $H(\sigma)$: The energy of configuration $\sigma$.
- $\sigma$: A vector representing the state of the spins (dimension $N$).
- $\sigma \in \{-1, -1\}^N$: Each spin takes a value of either $-1$ or $+1$, and the overall configuration forms an $N$-dimensional vector in spin space.
- $\Omega_{ij}$: Represents the strength of the coupling between spins $i$ and $j$.

This section explains the definition and calculation method of the evaluation metrics used in CIM-CACm. The evaluation metrics are listed in Table II below.

TABLE II. EVALUATION METTRICS [10]

| Metrics | Meaning | Value When Performance is Good |
|---|---|---|
| TTS (Time To Solution) | Product of time to solution and accuracy of the solution | Small |
| $p_0$ | Probability of finding the ground state | Large |

In addition, the evaluation metric TTS (Time to Solution) in [10] is defined by Equation (3).

$$TTS = \frac{\log(1-0.99)}{\log(1-p_0)} T \quad (3)$$

### D. Experimental Settings

#### 1) Overview

In this performance evaluation, we target the Wishart Planted Instances provided in the CIM-CACm benchmark [10]. This benchmark applies CIM-CACm to solve instances from the Wishart Planted Ensemble (WPE), a class of problems that can be systematically generated to control the difficulty of recovering the planted solution [10].

The benchmark includes problem sizes of $N$=60, 100, 140, and 200. In our experiments, we assume a use case where the goal is to improve upon already known high-performing parameter settings. Therefore, the initial parameter set $D$ is provided by the user and contains the best-known hyperparameters. We evaluate how much the proposed Method A and Method B can improve the performance compared to this baseline.

The best-known hyperparameters used in this study were obtained using BOHB [11], a hyperparameter optimization algorithm known for its efficiency among Bayesian optimization methods. These parameters are listed in Table III, and are used as the initial values for both Method A and Method B.

TABLE III. KNOWN BEST PARAMETERS (INITIAL VALUES)

| $\beta_1$ | $\beta_2$ | $\alpha$ | $\gamma$ | $\xi$ |
|---|---|---|---|---|
| 1.185 | 1.185 | 0.170 | 1.270 | 0.070 |

#### 2) Settings for Conventional Methods

The baseline method assumed in this evaluation is simultaneous hyperparameter optimization using Bayesian optimization. Specifically, we define the performance of the baseline as the TTS obtained by parameter tuning using Optuna's TPEsampler, a standard Bayesian optimization approach. The detailed experimental settings for the baseline method are as follows:
- Problem size: $N$=60
- Number of time steps $T$=1000
- Time step size $\Delta$=0.5 (fixed)

Table IV lists the hyperparameters to be optimized along with their respective search ranges.

The number of trials for hyperparameter search in this study is set to 100 and 1000, respectively, to evaluate performance under both limited and extended search conditions.

TABLE IV. PARAMETERS AND SEARCH SPACE

| Parameters | $\beta_1$ | $\beta_2$ | $\alpha$ | $\gamma$ | $\xi$ |
|---|---|---|---|---|---|
| Search Space | 0.000 ∼ 2.000 | 0.000 ∼ 2.000 | 0.000 ∼ 0.300 | 0.000 ∼ 2.000 | 0.000 ∼ 0.300 |

*3) Settings for Method A and Method B*

The detailed settings for Method A and Method B are described below.

First, the problem size $N$, the number of time steps $T$, and the time step size $\Delta$ are the same as those used in the baseline method. The hyperparameters to be optimized and their corresponding search ranges are also identical to those used in the baseline. It is important to note that Method A is an algorithm that sequentially explores each hyperparameter in a user-defined order. Therefore, the search order of the hyperparameters can vary. In this evaluation, the order was fixed as follows: $\beta_1 \rightarrow \beta_2 \rightarrow \alpha \rightarrow \gamma \rightarrow \xi$.

For Method B, the number of trials allocated to the initial evaluation phase $Y$ was set to $T_{initial}=20$, and the remaining trials were set to $T_{remaining} = T - N \times T_{initial} = 800$. Note that the choice of $Y$ is itself a hyperparameter of the algorithm, and therefore, this evaluation does not fully capture the performance potential of Method B.

In this study, the total number of hyperparameter optimization trials was set to 100 and 1000.

The following five search algorithms were used in the experiments:
1. Tree-structured Parzen Estimator (TPE)
2. Gaussian Process (GP)
3. Covariance Matrix Adaptation Evolution Strategy (CMA-ES)
4. Random Search
5. Grid Search

Both Method A and Method B are capable of dynamically switching between the above algorithms for each hyperparameter search. However, in this evaluation, dynamic switching was disabled, and a single search algorithm was applied throughout each experiment.

*E. Experimental Environments*

In this experiment, we used the Supercomputer "Flow" Cloud System [12] (hereafter referred to as the Flow Cloud), a supercomputing system installed at the Information Technology Center, Nagoya University.

The hardware configuration is summarized as follows. The CPU used is an Intel Xeon Gold 6230 with 20 cores, operating at 2.10–3.90 GHz, across 4 sockets. The memory is DDR4 2933 MHz, totaling 384 GiB (16 GiB × 6 modules × 4 sockets). The software versions are as follows: Python 3.7.6, numpy 1.19.0, pandas 1.1.5, matplotlib 3.3.4, torch 1.10.2, CACm 1.22, and optuna 4.0.0.

*F. Result*

*1) Comparison to Conventional Method and Proposal Methods*

The experimental results for Method A and Method B are presented below. Table V shows the results with 100 hyperparameter search trials, while Table VI presents the results for 1000 trials.

In both tables, the values represent the Time to Solution (TTS), where lower values indicate better performance. The search algorithms—TPE, GP, CMA-ES, Random, and Grid—were initialized with the best-known parameters. The TTS obtained using the best-known parameters, as shown in Table III, was 5926. Each method was evaluated through more than 10 independent runs, and the average TTS values are reported.

From Tables V and VI, we observe that Method B achieved the best performance with TPE, outperforming the other search algorithms. Moreover, all algorithms outperformed the baseline results obtained with the best-known hyperparameters. Both Method A and Method B were able to discover better parameter configurations than the previously best-known ones.

Additional comparisons, including speedup factors from the baseline method—i.e., the conventional approach involving simultaneous optimization of five hyperparameters using a fixed algorithm—are also discussed.

TABLE V. COMPARISON OF TTS BETWEEN CONVENTIOANL METHOD AND METHOD A AND METHOD B (NUMBER OF TRIALS: 100) TTS IS THE AVERAGE VALUE OVER 10 RUNS.

| Methods | TPE | GP | CMA-ES | Random | Grid |
|---|---|---|---|---|---|
| Conventional | 9277 | 21418 | 6884 | 9767 | 39641 |
| A | 4136 | 4896 | 5564 | 4448 | 4388 |
| B | **4048** | 4268 | 4959 | 4717 | 4091 |

TABLE VI. COMPARISON OF TTS BETWEEN CONVENTIOANL METHOD AND METHOD A AND METHOD B (NUMBER OF TRIALS: 1000) TTS IS THE AVERAGE VALUE OVER 10 RUNS.

| Methods | TPE | GP | CMA-ES | Random | Grid |
|---|---|---|---|---|---|
| Conventional | 9812 | 27191 | **675** | 3384 | 14160 |
| A | 4045 | 4140 | 4084 | 4134 | 4133 |
| B | 3600 | 3783 | 3710 | 3733 | 3755 |

TABLE VII. SPEEDUP FACTORS TO CONVENTIOANL METHOD (NUMBER OF TRIALS: 100)

| Methods | TPE | GP | CMA-ES | Random | Grid |
|---|---|---|---|---|---|
| Conventional | 1.00x | 1.00x | 1.00x | 1.00x | 1.00x |
| A | **1.43x** | 1.21x | 1.07x | 1.33x | 1.35x |
| B | **1.46x** | 1.39x | 1.19x | 1.26x | 1.45x |

TABLE VIII. SPEEDUP FACTORS TO CONVENTIOANL METHOD (NUMBER OF TRIALS: 1000)

| Methods | TPE | GP | CMA-ES | Random | Grid |
|---|---|---|---|---|---|
| Conventional | 1.00x | 1.00x | 1.00x | 1.00x | 1.00x |
| A | **1.47x** | 1.43x | 1.45x | 1.43x | 1.43x |
| B | **1.65x** | 1.57x | 1.60x | 1.59x | 1.56x |

In Table V and Table VI, method A and method B are:
- Method A is a sequential search based on a user-defined order
- Method B is a sequential search based on an automatically determined order.

Furthermore, using the values in Table V and Table VI, we compute the speedup ratios of Method A and Method B relative to the best-known parameter set, defined as:

$$Speedup = \frac{\text{Execution Time for Convenional Method}}{\text{Execution Time for Proposal Method}} \quad (4)$$

In Table VII, the maximum speedup achieved by Method A was confirmed to be 1.43x, while Method B achieved a maximum speedup of 1.46x. On the other hand, in Table VIII, Method A achieved a maximum speedup of 1.47x, and Method B achieved a maximum of 1.65x.

*2) Variance in Time to Solution (TTS)*

Fig. 4 shows the variance of TTS (based on 10 runs) for each algorithm selection under the conventional method, Method A, and Method B with 1000 trials.

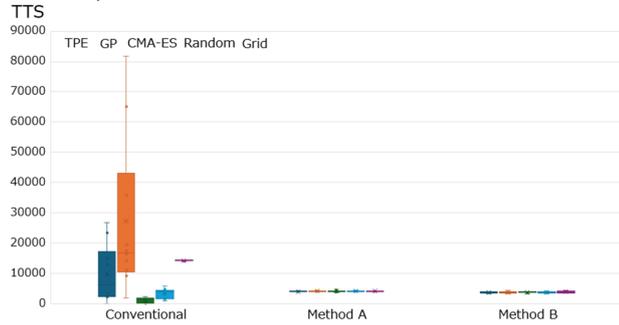

(a) Conventional Methods, Method A, and Method B.

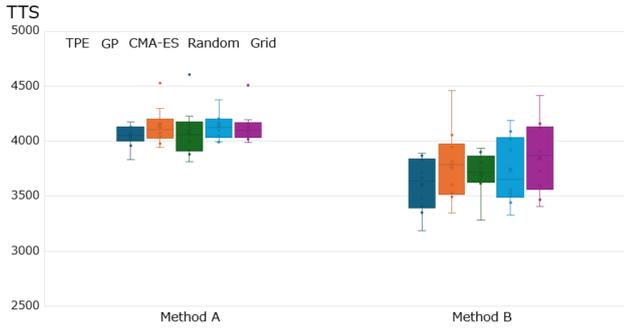

(b) Method A and Method B.

Fig. 4. Variance of TTS (based on 10 runs) for each algorithm selection under the conventional method, Method A, and Method B with 1000 trials.

Fig. 4 (a) shows that in the conventional method, the variance of TTS across measurements is significantly larger compared to Method A and Method B, depending on the specified algorithm. On the other hand, Fig. 4 (b) indicates that Method B generally achieves lower TTS values than Method A.

These results suggest that Method B is a more effective search method.

*3) Discussion*

In Tables V and VI, for the Bayesian optimization-based search methods TPE and GP, the proposed approach outperformed the conventional method—which simultaneously optimizes five hyperparameters—in both TTS and computational speed. This improvement is likely due to the nature of the target problem, where some hyperparameters have a minimal impact on TTS. Consequently, the proposed strategy of optimizing hyperparameters individually proved more effective. Nonetheless, analyzing the interdependencies among hyperparameters remains an open challenge for future research.

In the current evaluation, we also observed instances where the proposed methods were less effective when using CMA-ES as the search algorithm. One possible explanation is that CMA-ES, as an evolutionary algorithm, is well-suited to high-dimensional search spaces and leverages a covariance matrix to capture interdependencies among parameters during the search process. By optimizing hyperparameters independently—as done in the proposed methods—these interdependencies are disregarded, which may reduce search efficiency and hinder CMA-ES from reaching its full potential.

Another contributing factor to CMA-ES's reduced effectiveness in the proposed methods may lie in the parameter range settings used in Optuna. These ranges were defined based on preliminary evaluations and are likely centered near optimal values, allowing the baseline CMA-ES to perform well within a relatively constrained space. Supporting this, we observed that even the baseline Random Search—which typically underperforms compared to TPE—achieved better results than the baseline TPE in certain cases.

It is important to note that the above interpretations are based on exploratory observations, and no formal validation of these hypotheses has been conducted. Testing and verifying these hypotheses remain an important direction for future work.

## V. RELATED WORK

The following summarizes the positioning of this study in comparison with previous research.

1. Application of Early Stopping

Hyperparameter optimization methods such as BOHB [11] include an Early Stopping mechanism to eliminate unpromising trials. Optuna, the framework used in this study, also provides efficient pruning mechanisms. Specifically, it includes the Successive Halving Pruner, which implements the Successive Halving algorithm (SHA) [13], and the Hyperband Pruner, which extends SHA [14]. Although the impact of Early Stopping was not explicitly evaluated in this study, both Method A and Method B are compatible with the application of such techniques.

2. Reduction of Target Hyperparameters

The proposed Methods A and B are designed to improve the efficiency of Bayesian optimization by reducing the number of hyperparameters optimized simultaneously, thereby shrinking the search space. In contrast, this feature has not been considered in prior works such as [7] and [11].

3. Switching of Search Algorithms

The proposed methods offer flexibility in selecting the search algorithm dynamically during execution. If an appropriate search method is known in advance for a given hyperparameter, the algorithm can switch to it accordingly. In contrast, prior methods such as [7] and [11] do not incorporate runtime switching of optimization algorithms as a standard feature.

## VI. CONCLUSION

In this study, we proposed novel methods to address the limitations of Bayesian optimization when simultaneously tuning many hyperparameters that influence the performance of the CIM-CAC algorithm, which integrates CAC into the Coherent Ising Machine (CIM). The effectiveness of the proposed methods was evaluated by comparing their performance with that of conventional approaches.

A major challenge in applying conventional Bayesian optimization lies in the diminished performance gains when the number of simultaneously tuned hyperparameters is large—even as the number of search trials increases. Although Bayesian optimization improves efficiency by prioritizing unexplored regions, it remains susceptible to the curse of dimensionality under high-dimensional conditions. To overcome this issue, we introduced Method A, which restricts the number of hyperparameters optimized at any given time to one, allowing the optimization to proceed sequentially. To further enhance this strategy, we developed Method B, which dynamically determines the order of parameter optimization based on an initial evaluation phase.

Performance evaluation results demonstrate that when the total number of trials was limited to 100, Method A achieved up to a 1.43x improvement, and Method B up to a 1.46x improvement, in Time to Solution (TTS) compared to the baseline using the best-known hyperparameter settings. When the number of trials was increased to 1000, Method A achieved up to a 1.47x improvement, and Method B up to 1.65x. In all tested scenarios, Method B consistently outperformed Method A. These results confirm the effectiveness of the proposed methods in reducing the hyperparameter search space and improving the efficiency of Bayesian optimization by limiting the number of parameters optimized simultaneously.

Despite these promising outcomes, several challenges remain for both Method A and Method B.

First, with respect to Objective (1)—reducing the number of simultaneously optimized hyperparameters—Method A was evaluated by limiting the number of active parameters to one. However, in scenarios where strong interdependencies exist among multiple parameters, this strict reduction can result in inefficient search behavior. To address this, it is necessary to develop and evaluate techniques that analyze parameter dependencies and automatically determine an appropriate subset of hyperparameters to optimize jointly. For instance, if certain parameters are found to be highly correlated, they should be optimized together.

Second, concerning Objective (2)—the ability to apply different search strategies for different hyperparameters—both Method A and Method B are, in principle, capable of dynamically selecting suitable optimization algorithms, as described in this study. These methods allow flexible switching of search algorithms for each target parameter, provided that appropriate methods are identified beforehand. However, this dynamic selection capability has not yet been evaluated in the current experiments. Implementing and assessing this feature in both methods remains an important area for future work.

Finally, the proposed methods can be implemented within existing frameworks, languages, and tools for software auto-tuning (AT) [15][16][17]. Integrating the proposed approach into AT systems represents a significant step toward broader generalization. Expanding these methods for compatibility with diverse AT platforms is a key challenge that should be addressed in future research.

ACKNOWLEDGMENT

This research was supported by financial assistance and technical support from NTT Research, to whom the authors express their sincere gratitude.

This work was also supported by the Joint Usage/Research Center for Interdisciplinary Large-scale Information Infrastructures (JHPCN) and the High-Performance Computing Infrastructure (HPCI) under project number jh250015.

In addition, this research was funded by JSPS KAKENHI Grants JP23K11126 and JP24K02945.

The authors would also like to express their deep appreciation to Mr. Ichiro Takahashi of Information Technology Center, and Mr. Makoto Morishita of Graduate School of Informatics, Nagoya University, for their collaboration and for providing the related data.